# Chalcogen assisted contact engineering: towards CMOS integration of Tungsten Diselenide Field Effect Transistors

Ansh[1], J Kumar[1], R K Mishra[2], S Raghavan[2] and Mayank Shrivastava[1]

[1]Department of Electronic Systems Engineering, Indian Institute of Science, Bangalore-560012, India
[2]Centre for Nano-Science and Engineering, Indian Institute of Science, Bangalore-560012, India

**Abstract**

**One of the major roadblocks for the establishment of 2D semiconductor technology for CMOS integrated circuits is lack of industry scalable doping techniques that lead to 2D FETs with comparable n-type and p-type behavior. Here we demonstrate a Chalcogen based technique to alter the surface of $WSe_2$ to realize enhanced ambipolar behavior along with a complete transition from n to p-type behavior of $WSe_2$ FETs. The technique involves dry chemistry between Chalcogen atom and TMDC surface which leads to surface states that cause improved hole and electron injection through the FETs. We propose such a technique for realization of all $WSe_2$ based CMOS integrated circuits and therefore unveil its potential towards technology.**

Application of 2D semiconductors in transistor technology has consistently challenged the anticipated decline of Moore's law[1] along with advancement of semiconductor industry in terms of speed and power consumption. This has been possible because of the remarkable properties exhibited by low dimensional semiconductors which are suitable for field-effect transistor (FET) operation. 2D Transition Metal Dichalchalcogenides (TMDs)[2] are "beyond Graphene" materials which have finite non-zero bandgap like Si. This property along with atomically thin surface of 2D TMDs makes them indispensible for future of semiconductor industry especially for logic applications. In order to establish a 2D material specific transistor technology for complementary (CMOS) logic applications, it is of prime importance that both n and p type FETs exist with comparable performance. Unlike for Si and other bulk semiconductors, for 2D materials, absence of a tunable and CMOS compatible doping technique is a bottleneck in unveiling full technological significance of 2D semiconductors[4]. The polarity of 2D TMD FET relies on the position at which the metal Fermi-level pins when contacted with the TMD surface[5]. Usually, surface defects in TMDs manifest themselves in bandstructure as energy states within the bandgap. For example, S vacancies in $MoS_2$ and $WS_2$ surfaces lead to bandgap states lying closer to the conduction band minimum (CBM)[5]. As a result, metal Fermi level is strongly pinned closer to the conduction band thereby facilitating electron conduction across the contact. This is a strong reason of the observed unipolar electron conduction or n-type behavior in $MoS_2$ and $WS_2$ FETs irrespective of the metal work function[6]. Unlike $MoS_2$ and $WS_2$, $MoSe_2$ and $WSe_2$ FETs exhibit ambipolar transistor behavior which is attributed to two different phenomena- smaller bandgap and FLP at the midgap energy, observed in these TMDs ($MoSe_2$ and $WSe_2$) respectively[7]. The ratio of electron current and hole current in these TMDs depend on the layer thickness and therefore it becomes even more difficult to realize individual n-type and p-type devices on monolayers owing to their larger bandgap. For the development of 2D semiconductor based CMOS circuit building blocks, realization of unipolar n-type and p-type FETs on the same material is preferred over ambipolar FETs so that device metrics like leakage current, $V_T$ and $I_{ON}/I_{OFF}$ do not deteriorate circuit metrics like noise margin and static power loss. As discussed earlier, in order to establish a 2D semiconductor specific technology for complementary logic applications, demonstrating transistors with comparable n-type and p-type performance is vital. Theoretical studies have been performed to study the effect of various possible dopants for 2D materials so that a control on their polarity can be achieved[8, 9, 10]. However, experimental demonstrations achieve polarity control in 2D TMD FETs are rare[11, 12]. Few techniques involve doping during the growth process and others involve wet chemistry that leads to surface charge transfer to facilitate increase in specific carrier (electron or hole) concentration. Although these techniques have successfully realized the desired doping and or polarity in FETs, they are not industry scalable and CMOS process compatible. In this work, we introduce a unique Sulfur based scalable and CMOS process compatible technique to cause a polarity reversal from n-type to p-type conduction in $WSe_2$ FETs. Subsequent theoretical investigation reveals the physical phenomenon occurring at the surface thereby enabling us to propose a process flow for the fabrication of $WSe_2$ based CMOS integrated circuits.

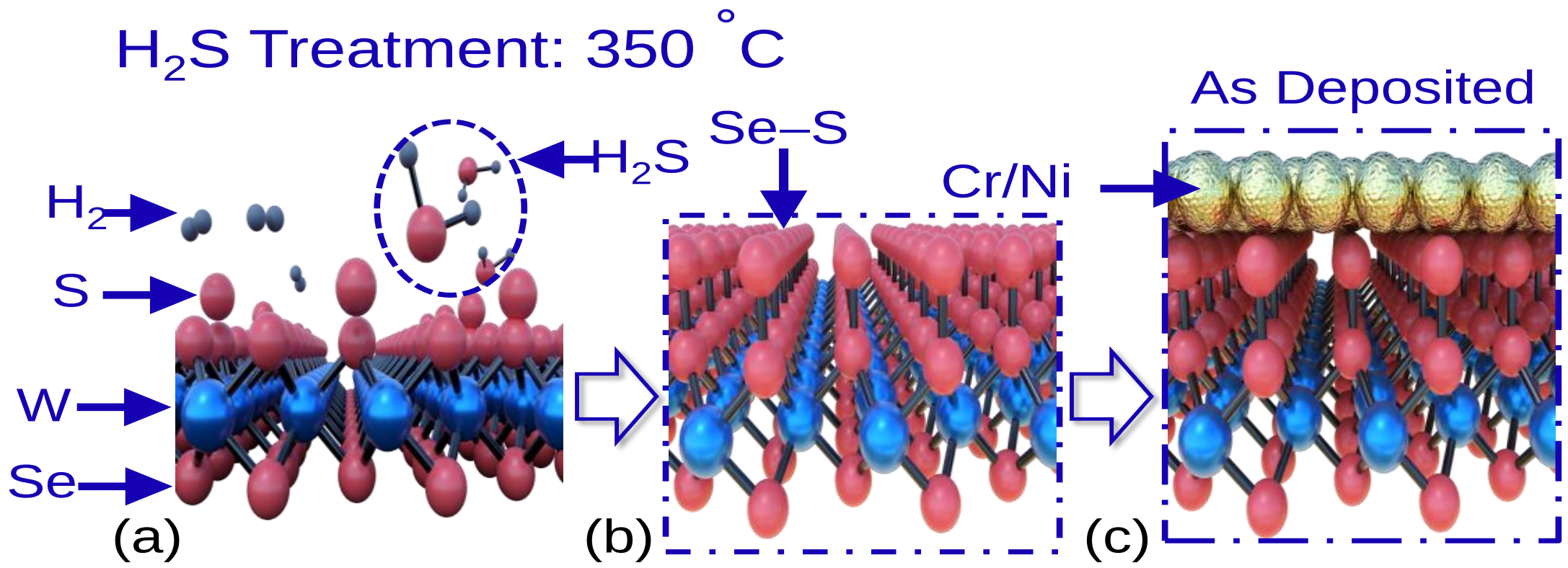

Figure 1: Partial decomposition of $H_2S$ on TMD surface leaves behind S species on the surface [13]. The S atom bonds with Se atom of TMDs and $H_2$ is liberated out, as depicted in (i) & (ii)

As discussed earlier, Chalcogen vacancies are responsible for the observed polarity in TMD FETs. It is interesting to visualize the effect of reduced Chalcogen vacancy on the device behavior. The situation becomes more interesting when Sulfur (S) is visualized on the surface of a non-Sulfur TMD like $WSe_2$. We hypothesize that smaller size of S atoms (which means presence of lower energy orbitals compared to that of Se) should essentially influence the position at which the metal Fermi level pins. To observe its effect, $WSe_2$ samples were treated with $H_2S$ gas which is widely used as the source of S during Chemical vapor deposition (CVD) growth process of S based TMDs. The primary reason for the use of $H_2S$ is that it decomposes on the surface of transition metal and TMDs at high temperatures to give S. Studies[13] have shown that at lower temperatures it tends to partially decompose and a catalytic step is required to completely decompose it. Here we adapted a process where $WSe_2$ samples are exposed to $H_2S$ at 350 °C. The mechanism is shown in figure 1 and it is expected that S is incorporated on the surface of $WSe_2$ upon exposure to $H_2S$ at specified conditions. In order to ensure that such a treatment does not drastically change the fundamental molecular structure of $WSe_2$, Raman spectra is captured before and after $H_2S$ exposure, as shown in figure 2. A blue shift is observed in the Raman spectra after exposure which corresponds to enhanced electron-phonon interaction[14]. Moreover, presence of the signature $E_{1g}$ and $A_{2g}$ peaks imply that the fundamental molecular structure that causes the corresponding Raman active modes in $WSe_2$ is intact, post exposure. Presence of S on the surface of a non-S TMD can be easily identified through elemental analysis of its surface. We compare the X-ray Photoluminescence Spectroscopy (XPS) spectra of $WSe_2$samples before and after $H_2S$ exposure. Interestingly, comparison of XPS spectra reveals the anticipated presence of S on the surface after $H_2S$ exposure. As shown in figure 3, S 2p peaks are observed in the XPS spectra of exposed flakes which were otherwise missing. Other observations from the XPS data are lowering of shoulders around W-5p and 4p peaks along with those around Se-3d peaks. Lowering of shoulders close to the peaks is often attributed to lower defect density.

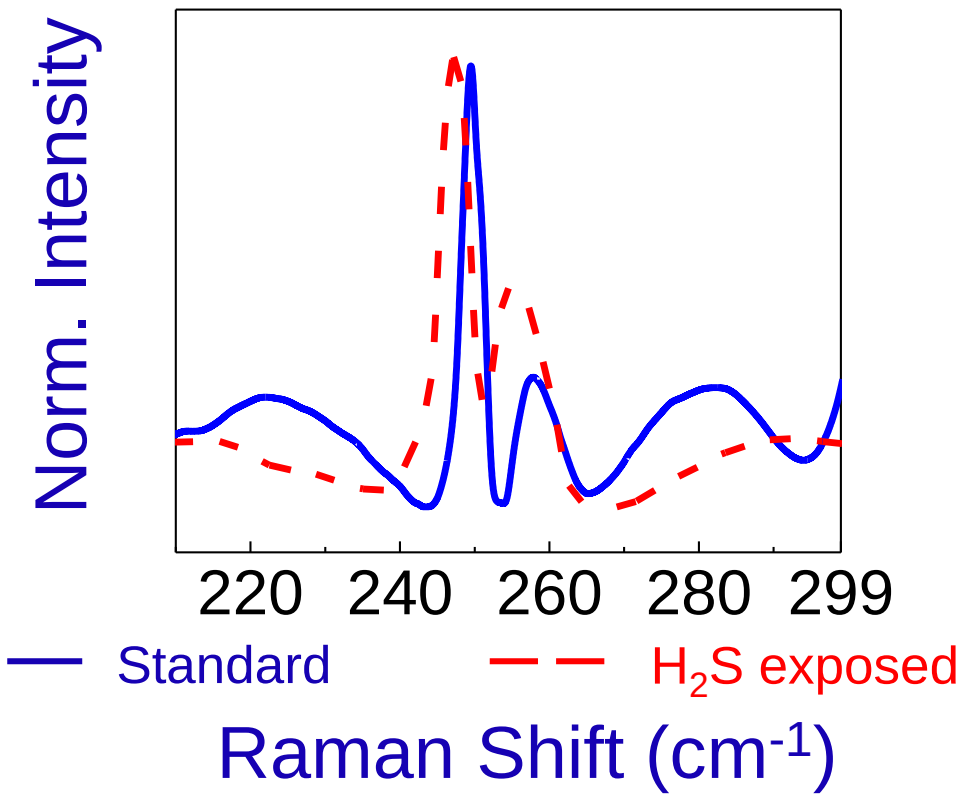

Fig. 2: Raman spectra of as-exfoliated and engineered ($H_2S$ treated) $WSe_2$ flakes. Narrowing of the Raman peaks along with lowering of shoulders close to peaks depict satisfaction of Surface defects. A blue shift is observed implying increased electron-Phonon coupling.

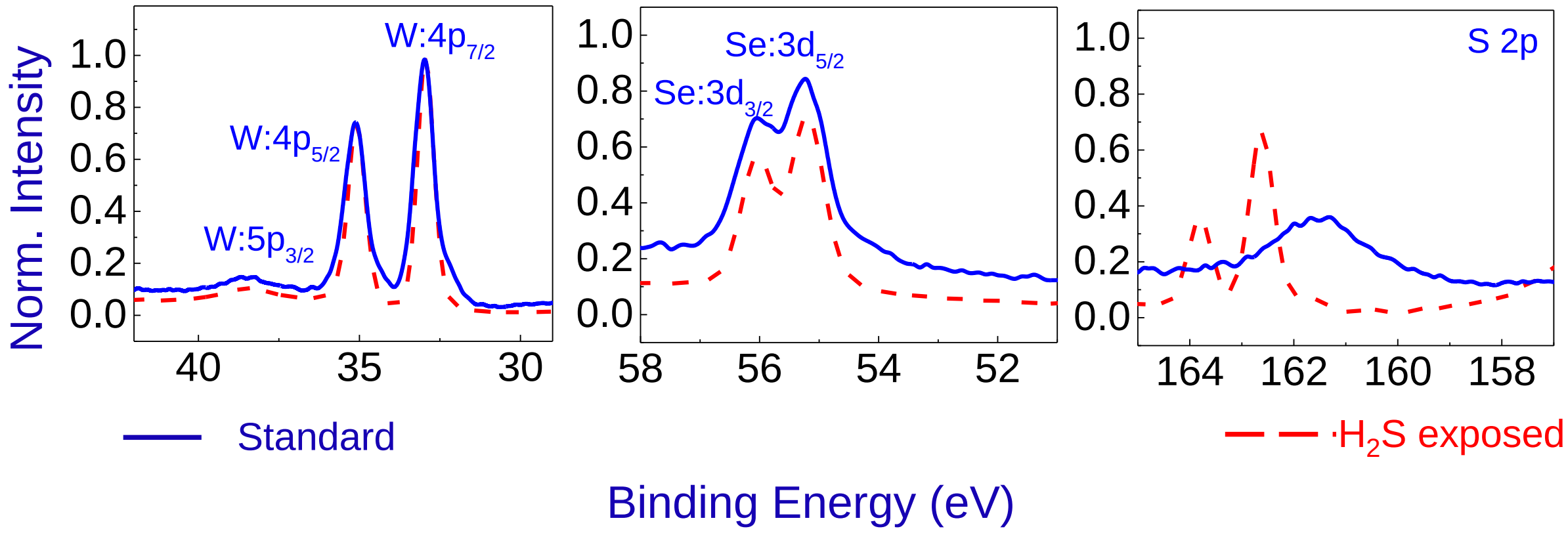

Fig. 3: XPS spectra for $WSe_2$ sample before and after $H_2S$ exposure show reduced shoulders near 4p peaks of $WSe_2$ which implies reduction in Se vacancy/defects and W dangling bonds present in the material. Presence of S in engineered $WSe_2$ surface reveals that $H_2S$ exposure is a reliable method to introduce S atoms at the surface to introduce acceptor states in the bandgap that can in turn enable FLP near VBM.

Impact of S at the $WSe_2$-metal interface (engineered contact) is explored by fabricating FETs using the standard scotch tape method. $WSe_2$ (from 2D Semiconductors) flakes are exfoliated on top of a 90 nm thermally grown $SiO_2$ on $p^{++}$ doped Si substrate. Electron beam lithography is used to pattern source/drain (S/D) contact regions which are subsequently exposed inside an electron beam evaporator to deposit the desired metal stack(Ni/Au or Cr/Au) of thickness 5/50 nm. Lift-off followed by 200°C thermal anneal step is performed to complete the fabrication of standard (without $H_2S$ exposure) back gated $WSe_2$ FETs with $L_{ch}$ = 1µm. During the fabrication process, the channel region was masked by 15nm $Al_2O_3$ for both standard and contact engineered devices. These FETs are then electrically characterized inside a vacuum probe station at room temperature. In order to quantify the effect of $H_2S$ exposure on the device behavior, data is compared before and after $H_2S$ exposure. It is worth mentioning that the same uniformly thick flake is used to fabricate contact engineered (exposed to $H_2S$) FETs by patterning S/D regions on the un-processed portion of the same flake that is used to fabricate standard FETs. This is done to eliminate variability due to varying flake thicknesses across different flakes. Next the samples are loaded inside a furnace kept at 350 °C with a $H_2S$ partial pressure of 20 torr after which the contact engineered FETs are fabricated by following the exact same process as discussed earlier. It is observed that standard FETs with Ni contacts exhibit N-type behavior which is consistent with previous reports[xx], as shown in figure 5(a). Upon $H_2S$ exposure, these device exhibit enhanced ambipolar behavior along with improved electron and hole injection across the device, as shown in figure 5(b). Cr is a low work function metal and therefore standard $WSe_2$ FETs with Cr contacts are expected to exhibit dominant N-type behavior with lower electron injection efficiency, which has been experimentally realized and shown in figure 5(c). Absence of hole conduction is observed at higher negative gate voltages is improves by 100 times after $H_2S$ exposure. It is observed, in figure 5(d), that contact engineered Cr devices exhibit p-type behavior in contrast to standard Cr devices with a negligible change in electron current. $H_2S$ exposure has resulted in huge positive shift in threshold voltage ($V_T$) over all devices irrespective of contact metal. A positive shift in $V_T$ indicates that the channel remains depleted for a wider range of gate voltages. This observation along with improved hole conduction suggests strong FLP toward the VBM of $WSe_2$ after $H_2S$ exposure. In order to understand the effect of S on the bandstructure of $WSe_2$, atomistic calculations are performed to calculate the bandstructure and Density of states (DOS) for three different surface topologies: defect-less $WSe_2$, $WSe_2$ with Se vacancy and $WSe_2$ with S at interstitial site at the surface.

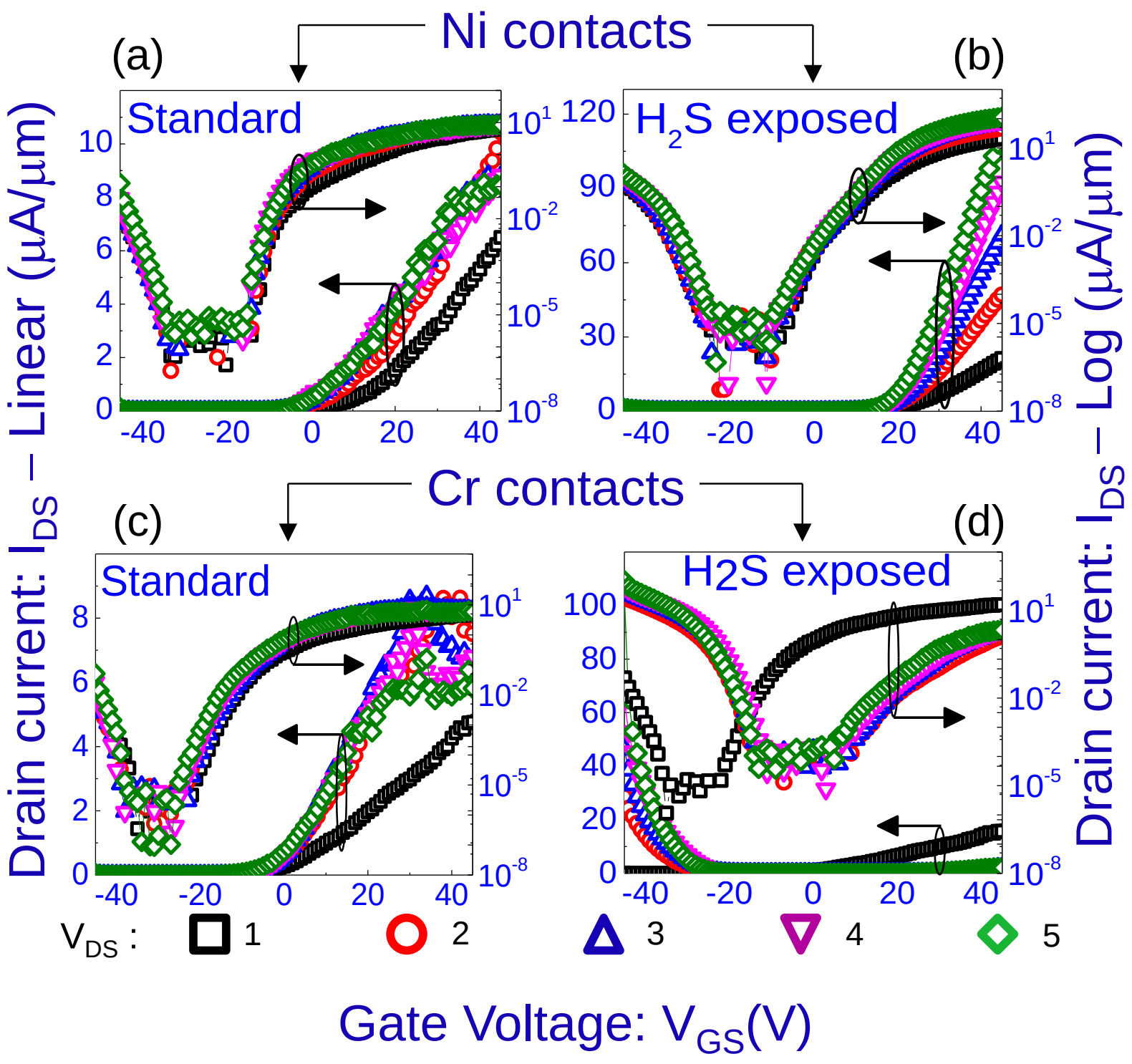

Figure 5: Transfer characteristics of Ni (a,c) and Cr (b,d) contacted FETs before and after $H_2S$ assisted contact engineering.

Figure 6 signifies DOS for the three surface topologies. It is observed that when S occupies an interstitial site on the surface of $WSe_2$, shallow states with large DOS are introduced near the VBM (acceptor states). This is in contrast to the presence donor states near the CBM when Se vacancy is present on the surface. On the basis of theoretical and experimental observations of presence of acceptor states and enhanced electron and (or) hole conduction for different contact metals, we theorize that presence of S on the surface of $WSe_2$ alters the interface in two ways- (i) introduces states at relatively lower energies that pin the metal Fermi level closer to the VBM; (ii) bonding between metal atom and TMD at the interface improves or remains unaffected (depending on the contact metal) that influences both kind of carrier injection into the device.

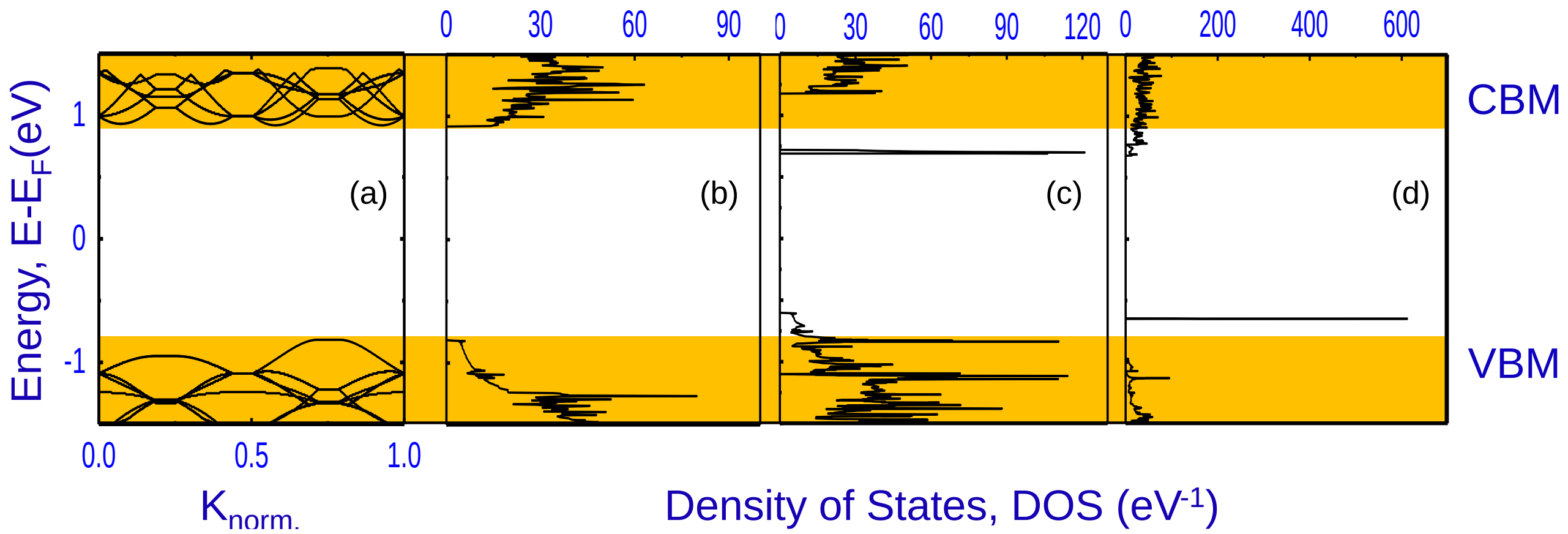

Figure 6: (a) Bandstructure of monolayer $WSe_2$ calculated in ATK. (b) Density of states calculated at different energies for (b) defect-less $WSe_2$, (c) $WSe_2$ with Se vacancy and (d) $WSe_2$ with S at interstitial site on the surface. It is observed that acceptor states are introduced in the bandgap when S stays at an interstitial site due to which FLP at lower energies is expected upon contact with metal unlike the case with Se vacancy where donor states are present in the bandgap.

Bonding at the interface manifests as Schottky Barrier Width (SBW) and FLP near VBM leads to lower Schottky Barrier Height (SBH) for holes (conversely, higher SBH for electrons). Lower SBW improves tunneling probability at the contact, on the other hand, lower SBH improves thermionic emission. Through output characteristics and electronic band theory, the effect of $H_2S$ exposure on switching mechanism of $WSe_2$ FETs with Ni and Cr is explained in figure 7, 8 and 9, 10 respectively. In standard Ni devices, sub-micron current at $V_{GS}$ = -45 V and negligible current at relatively positive $V_{GS}$ until at $V_{GS}$ = -5V where significant current is observed signifies the ambipolar behavior of the FET with onset of hole conduction at $V_{GS}$= -45 V and that for electrons is $V_{GS}$= -5 V, as shown in figure 7(a). The electron current is further enhanced for positive gate voltages beyond $V_{GS}$= -5 V, figure 7(b), which is a result of the device in strong inversion regime. Effect of gate voltage on the electronic bands of a 2D semiconductor and the switching mechanism of standard Ni-FET is discussed in figure 7 (c)-(g). Due to the gate field effect, the bands bend downwards for positive gate voltages thereby lowering the SBW and SBH to allow electron injection through improved tunneling (blue arrows) and thermionic emission (red arrows) respectively. As shown in figure 7 (e)-(g), electron tunneling increases as the gate voltage becomes more positive whereas the thermionic component remains almost constant. Very high negative gate field ($V_{GS}$ = -45 V) bends the bands so much that tunneling barriers for holes are formed thereby facilitating hole injection as shown in figure 7(c) resulting in ambipolarity.

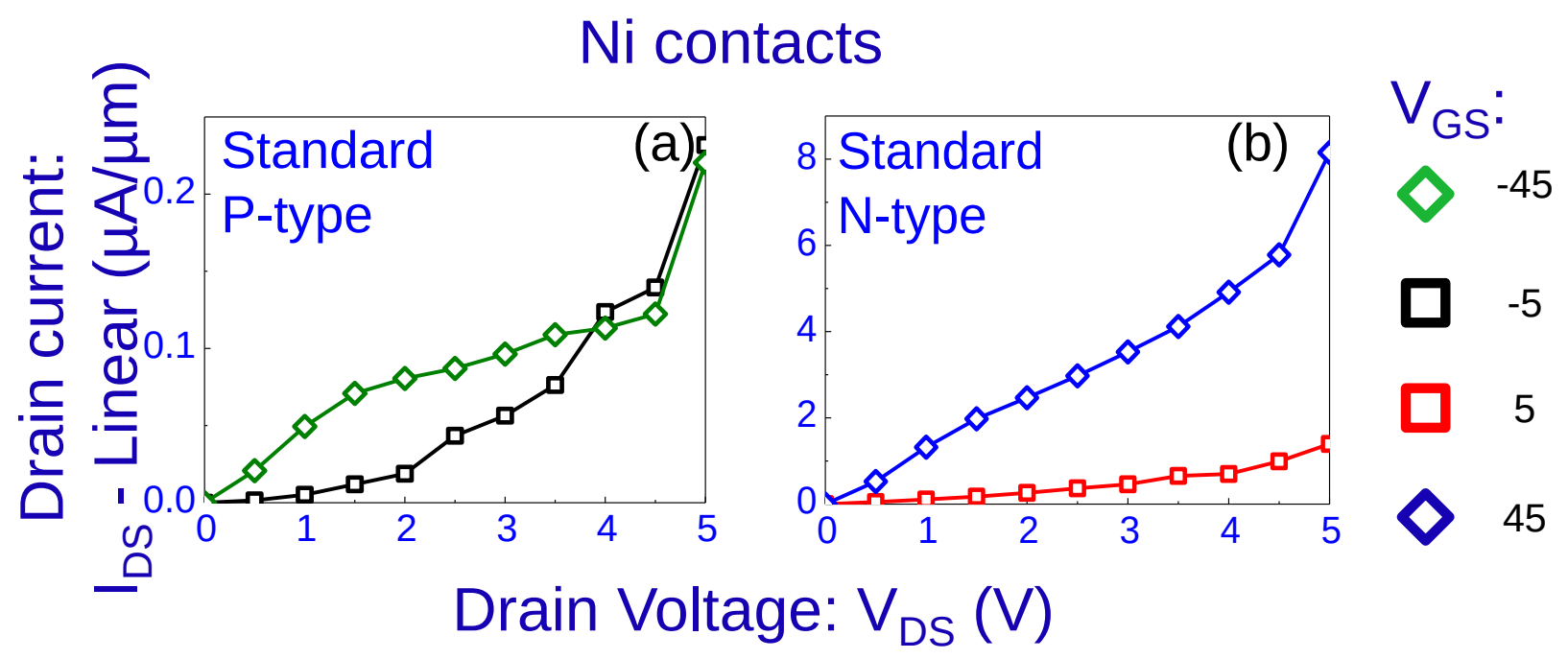

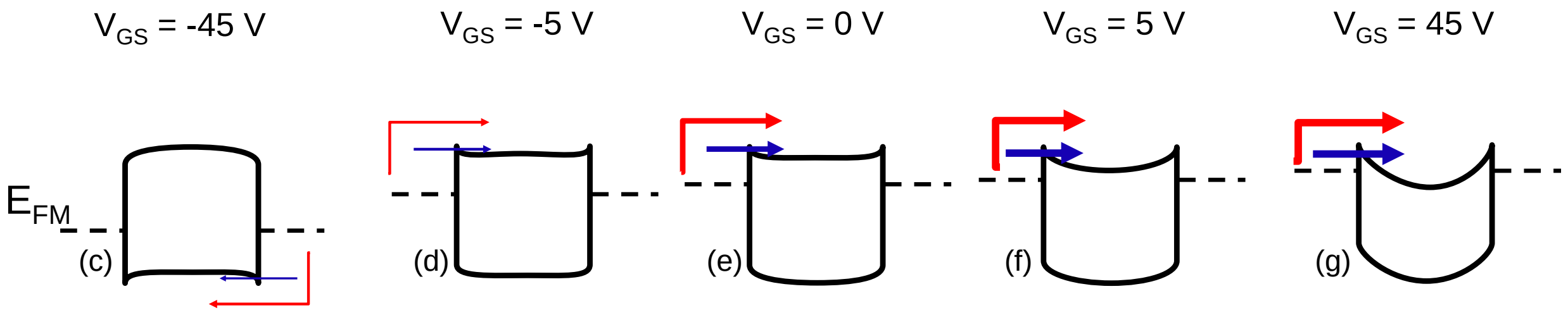

Figure 7: (a) p-type and (b) n-type behavior of standard Ni FETs along with the switching mechanism (c)-(g). Bands bend downwards at positive gate (f, g) voltages to reduce the SBW at the contacts thereby increasing the tunneling component of current (blue arrow) while the thermionic component is a function of drain voltage and remains constant for a certain drain voltage. As the gate voltage is tuned negative (c, d) bands start to move upwards. As a result, at sufficiently high negative gate voltage the Fermi-level aligns such that SB for holes is established and hole injection initiates.

$H_2S$ exposure, as validated by XPS and bandstructure calculations, introduces states near the VBM which is reflected as improved hole current at $V_{GS}$ = -45 V, in figure 8(a). Moreover, the onset of n-type conduction has shifted from $V_{GS}$ = -5 V to 20 V which is a direct implication of channel depletion at higher gate voltages (positive shift in $V_T$). Such a huge shift in $V_T$ is attributed to influence of S on the free electrons at the contact which bounds the electrons strongly as discussed earlier. Along with enhanced hole injection, electron injection has improved in Ni device after $H_2S$ exposure which is expected to be a result of improved bonding between $H_2S$ treated $WSe_2$ and Ni. This has been validated by Mulliken Charge Population (MCP) and bond length calculations included in the SI. As discussed earlier, improved bonding leads to smaller SBW thereby improving the tunneling current at the contact. From figure 8 (c)-(g), the switching mechanism of contact engineered FETs is discussed by showing narrower Schottky Barrier (SB) and improved tunneling current (thicker red arrows at higher gate voltage). So $H_2S$ exposure results in two different phenomena in Ni FETs- (i) FLP near VBM that leads to easier hole injection and (ii) improved hole and electron tunneling across the barriers because of narrower SBs.

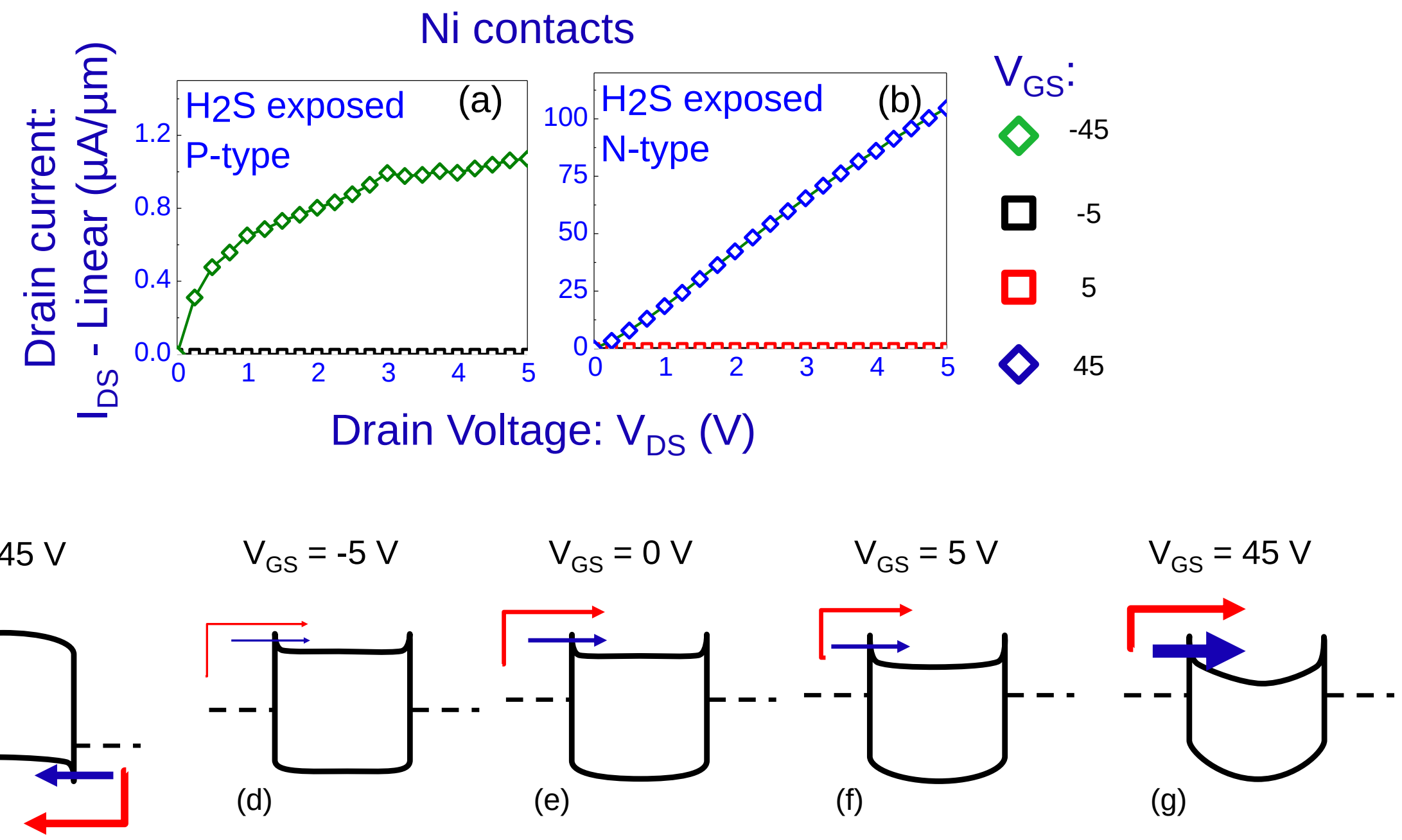

Figure 8: (a) p-type and (b) n-type behavior of contact engineered Ni FETs along with the switching mechanism (c)-(g). $H_2S$ exposure is expected to improve the bonding at the contact and narrow down the SB. Bands bend downwards at positive gate (f, g) voltages to further reduce the SBW at the contacts thereby increasing the already improved tunneling component of current (blue arrow) while the thermionic component is a function of drain voltage and remains almost constant for a certain drain voltage. As a result of $H_2S$ exposure, the Fermi-level aligns closer to the VBM allowing easy realization of SB fore holes Therefore, as the gate voltage is tuned negative (c, d), bands start to move upwards and enhanced hole injection is achieved.

The transistor switching mechanism for $WSe_2$/Cr FETs is shown in figure 9. It is clear that standard Cr devices exhibit dominant N-type behavior with significant current at $V_{GS}$ = -5 V compared to that at further negative voltages and relatively higher current at positive gate voltages. Effect of gate electric field on the energy bands of the channel and its implications on device switching behavior are shown in figure 9(c)-(g), which is similar to standard Ni contact FETs except that Cr devices exhibit lower hole and electron current. $H_2S$ exposure for contact engineering of Cr devices leads to P-type FETs as shown earlier in the transfer characteristics. Figure 10 (a)-(b) shows that contact engineered FETs exhibit significant current at $V_{GS}$ = -45 V which is absent in standard devices. Moreover, the onset of N-type conduction has shifted to higher gate voltage which leads to a huge shift in $V_T$ as shown in figure 6. Unlike for Ni FETs, the electron and hole current remain similar in Cr FETs for the same overdrive voltage, post-$H_2S$ exposure. This is attributed to zero or marginal reduction in SBW before and after exposure that results in similar hole and electron tunnel injection across the contacts while the thermionic component of current remains unchanged. As discussed earlier for Ni FETs, in contact engineered FETs, FLP occurs close to the VBM which increases SBH for electrons while decreasing SBH for holes due to which the thermionic emission of electrons is suppressed and that for holes improves. Unlike in Ni FETs, in Cr FETs only one phenomenon significantly alters the switching behavior- FLP near VBM due to acceptor states introduced by S atoms on the surface. This has been explained in figure 10 (c)-(g), where SBW has been shown to have reduced after $H_2S$ exposure that in turn results in enhanced tunnel current (thicker blue arrow) for holes (at $V_{GS}$ = -45 V) as well as electrons (at $V_{GS}$ = 20 to 45 V). Marginal or no improvement in bonding between $H_2S$ treated $WSe_2$ and Cr is attributed to half-filled d and s orbitals in Cr as discussed next. The primary reason for stronger bonding with Ni ([Ar] $3d^8$ $4s^2$) is its partially filled 3d orbitals as compared to Cr ([Ar] $3d^5$ $4s^1$) that has half-filled 3d and 4s orbitals. It is well known that a half filled electronic configuration is more stable than a partially filled electronic configuration and therefore, Ni is expected to exhibit better bonding properties than Cr thereby leading to narrower SB at the contact.

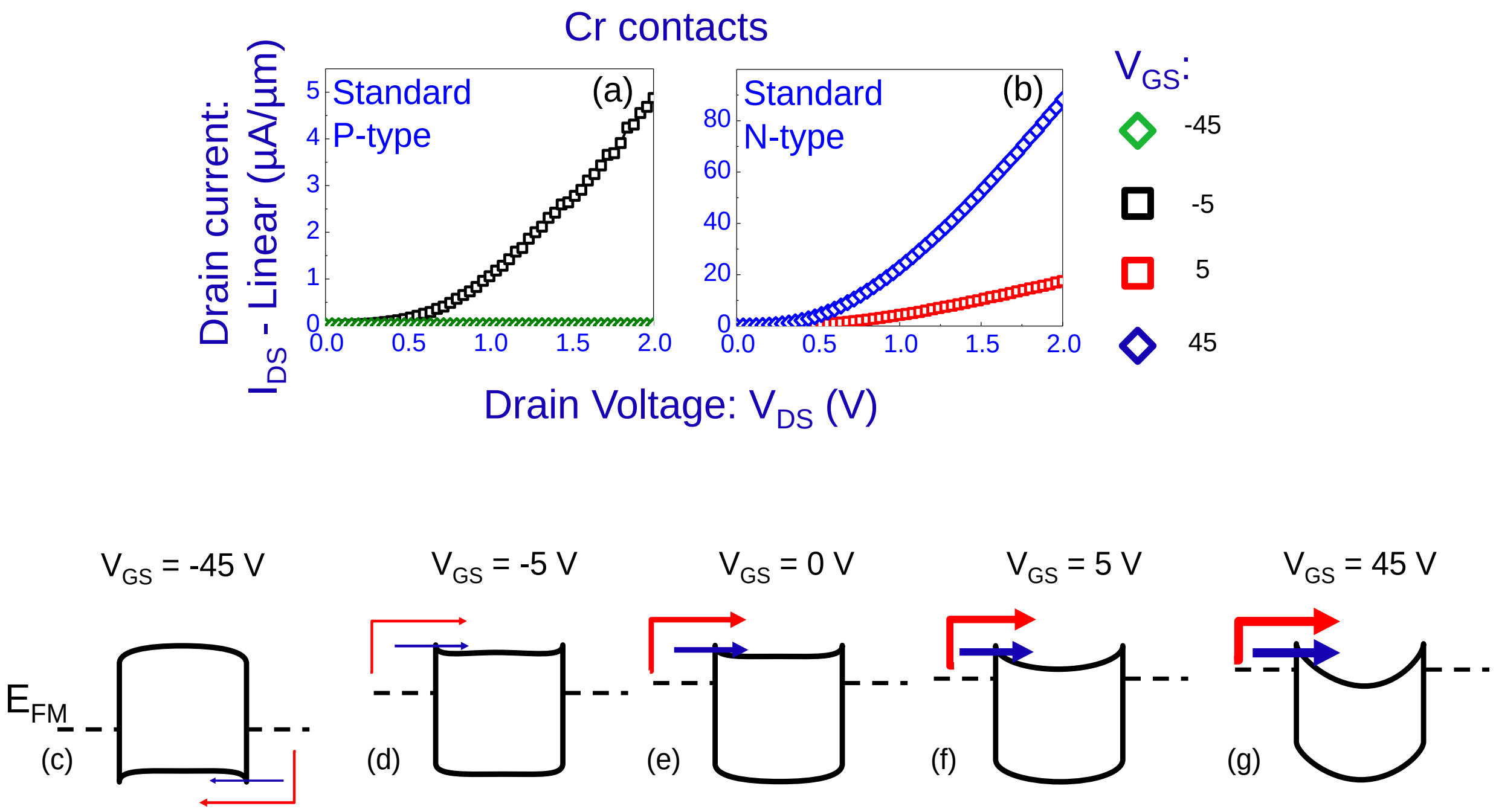

Figure 9: (a) p-type and (b) n-type behavior of standard Cr FETs along with the switching mechanism (c)-(g), similar to that of Ni FETs in figure 7.

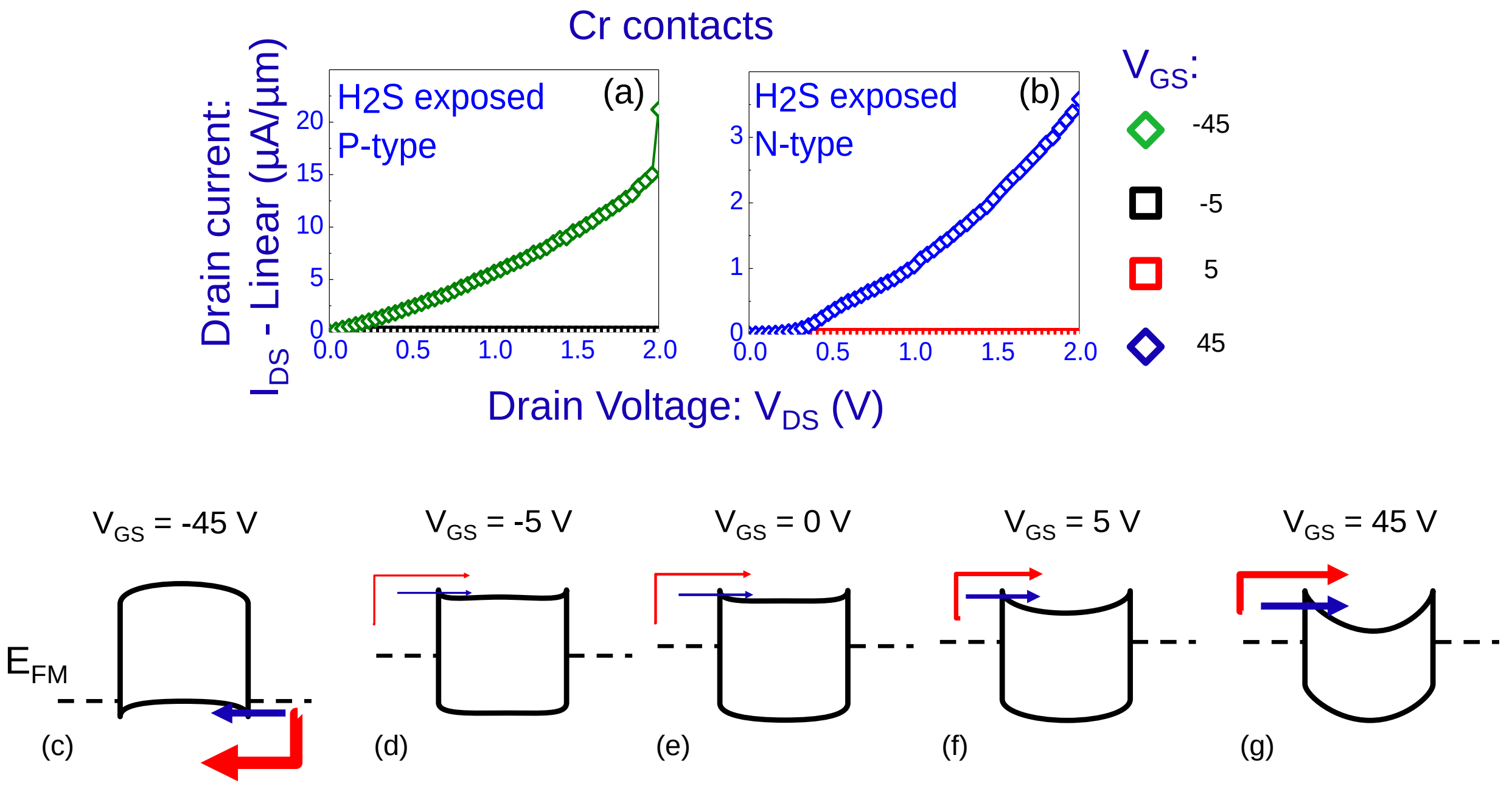

Figure 10: (a) p-type and (b) n-type behavior of contact engineered Ni FETs along with the switching mechanism (c)-(g). $H_2S$ exposure is expected to improve the bonding at the contact and narrow down the SB. Bands bend downwards at positive gate (f, g) voltages to further reduce the SBW at the contacts thereby increasing the already improved tunneling component of current (blue arrow) while the thermionic component is a function of drain voltage and remains almost constant for a certain drain voltage. As a result of $H_2S$ exposure, the Fermi-level aligns closer to the VBM allowing easy realization of SB fore holes  Therefore, as the gate voltage is tuned negative (c, d), bands start to move upwards and enhanced hole injection is achieved.

In order to quantify the effect of $H_2S$ on the contacts, contact resistance is extracted using Y-function method. Although this method over-estimates the value, it can be used to identify the impact by relating the $R_C$ values before and after the exposure. It is observed that, in figure 11, that $R_C$ for hole injection reduces drastically upon $H_2S$ assisted contact engineering for both Ni and Cr contacted devices. However, $R_C$ for electrons reduces only marginally in both cases, post treatment. Hole contact resistance is altered by two phenomena- FLP near VBM and improved bonding for Ni, as discussed earlier, which are responsible for such a huge improvement in its values post-exposure. For electrons, only in case of Ni the bonding is expected to improve which leads to slightly improved $R_C$ after exposure, unlike Cr where the bonding remains almost constant.

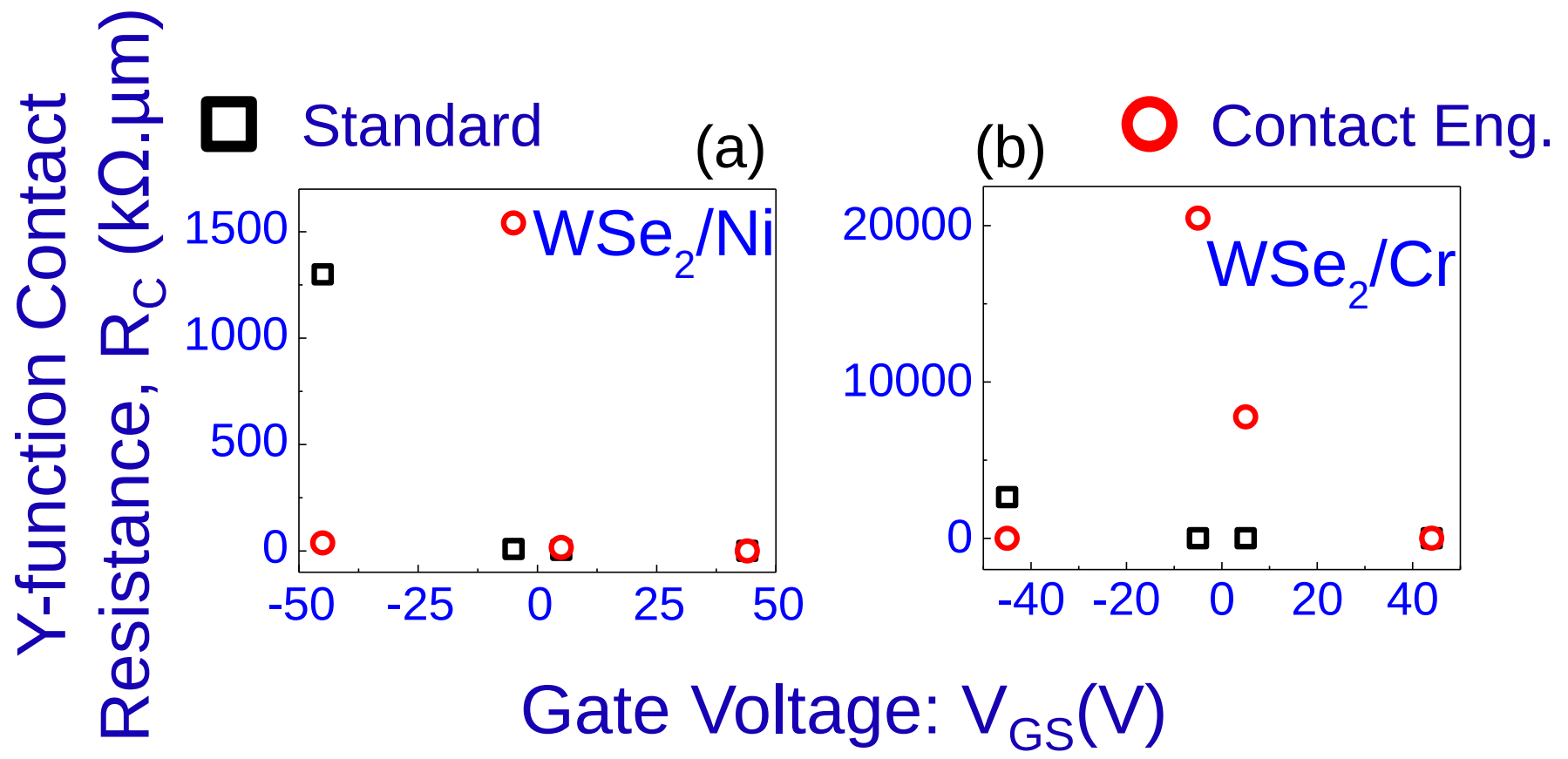

Figure 11: Y-function extracted $R_C$ of $WSe_2$ standard and contact engineered FETs with (a) Ni and (b) Cr contacts.

Finally, we propose a fabrication process of $WSe_2$ NMOS and PMOS FETs on the same semiconductor film for CMOS circuit integration in order to develop a more feasible $WSe_2$ based 2D transistor technology. The fabrication process, as shown in figure 12, includes fabrication of standard (NMOS) and contact engineered (PMOS) $WSe_2$ FETs along with a passivation layer that acts as mask for $H_2S$ exposure as well as the top gate dielectric.

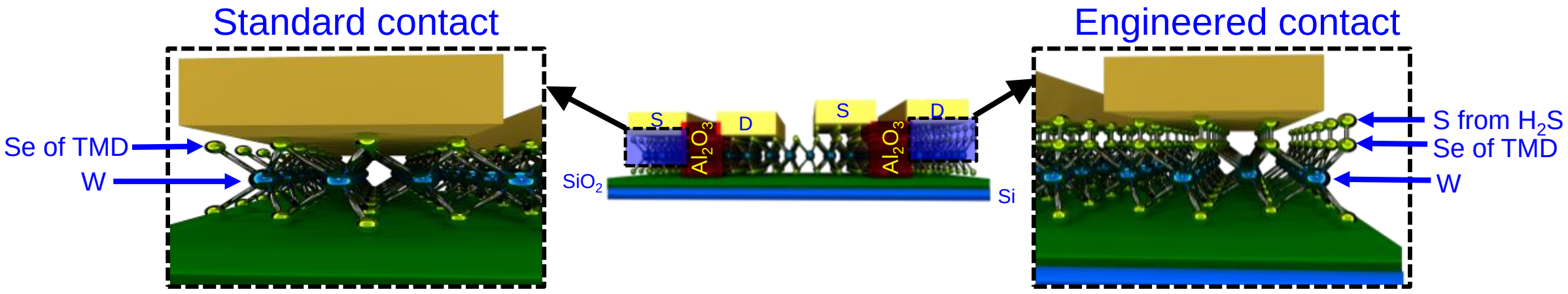

Figure 12: Scheme followed to fabricate standard and contact engineered FETs on the same TMD film on top of $SiO_2$/Si sample with $Al_2O_3$ passivation on the channel region.

In summary, we have demonstrated a Chalcogen based technique to improve ambipolarity and achieve polarity reversal in $WSe_2$ FETs. Physical insights into the role of S on the surface along with experimental observations suggest improvement in hole and (or) electron conduction through the channel by aligning the metal Fermi-level closer to the VBM and (or) improving bonding between $WSe_2$ and metal. It was observed that FETs with Ni contact exhibit stronger ambipolar behavior after $H_2S$ assisted contact engineering. On the other hand, in Cr contacted FETs, a transition from n-type to p-type behavior is achieved which can enable CMOS circuit integration of $WSe_2$ based NMOS and PMOS. This metal specific impact of $H_2S$ on $WSe_2$ arises from the fundamental difference between availability of atomic orbitals in different metals. These results clearly suggest the importance of this method towards 2D semiconductor complementary logic applications where both n-type and p-type device operation on the same integrated chip are desired.